\documentstyle[epsf]{article}
\begin{document}

\parskip 2mm plus 1mm \parindent=0pt
\def\cl{\centerline}
\def\hs1{\hskip1mm}\def\hr{\hskip.2mm} 
\def\h10{\hskip10mm}\def\hsp{\hskip.5mm} \def\page{\vfill\eject}
\def\hx{\hskip10mm\hbox} \def\vs{\vskip4mm}
\def\<{\langle} \def\>{\rangle}   
\def\d{{\rm d}} \def\de{\partial} \def\Tr {{\rm Tr}}
\def\dag{^\dagger}\def\N{{\cal N}}

\def\ne{=\hskip-3.3mm /\hskip3.3mm} 
\def\half{{\scriptstyle{1\over 2}}} \def\quter{{\scriptstyle{1\over 4}}} 
\def\Im{{\rm Im}} \def\Re{{\rm Re}} \def\I{{\rm I}} 

\def\be{\begin{equation}}\def\ee{\end{equation}}
\def\bl{\be\label{}}

\vs\cl{\bf A general computer program for the Bell detection loophole} 

\vs

\centerline {by}

\vs

\cl {Roberto M. Basoalto and Ian C. Percival}\vs 

\centerline {Department of Physics} 
\centerline {Queen Mary and Westfield College, University of London} 
\centerline {Mile End Road, London E1 4NS, England} 
\vskip5mm\cl{i.c.percival@qmw.ac.uk}
\vskip15mm \centerline{\bf Abstract} \rm 

The difference between ideal experiments to test Bell's weak
nonlocality and the real experiments leads to loopholes.  Ideal
experiments involve either inequalities (Bell) or equalities
(Greenberger, Horne and Zeilinger).  Every real experiment has its own
critical inequalities, which are almost all more complicated than the
corresponding ideal inequalities or equalities.  If one of these
critical inequalities is violated, then the detection loophole is
closed, with no further assumptions.  If all the critical inequalities
are satisfied, then it remains open, unless further assumptions are
made.  The computer program described here and published on the
website 

\cl{\it  http://www.strings.ph.qmw.ac.uk/QI/main.htm}

obtains the critical inequalities for any real
experiment, given the number of allowed settings of the angles and the
corresponding possible output signals for a single run.  Given all the
necessary conditional probabilities or rates, it tests whether all
these inequalities are satisfied.

\vskip55mm \vfill
\page

Weak nonlocality, or nonlocality in the sense of Bell, applies to
systems whose properties cannot be explained using local hidden
variables [1].  It is defined in [2] and in the next section in terms of
transfer functions for input-output systems.  The original Bell
inequalities and their generalizations lead to tests for weak
nonlocality in the laboratory.

There have been many experiments designed to do this.  It is
thirty-six years since Bell obtained the first inequality and
tentatively suggested an experiment, and about three decades since the
first experiments.  Yet there is still no published unambiguous
experimental demonstration of weak nonlocality, because of the
detection loophole [3,4,5,6,7,8], which we now describe.

Real experiments have outputs that are excluded in ideal experiments.
For example, in the original ideal Bell experiment, two entangled
particles are detected for every run.  But in a real experiment, because
of imperfect detectors, or losses due to absorption or imperfect
collimation, only one of these particles may be detected.  Such outputs,
that are present in the real experiment, but not in the ideal experiment,
affect the inequalities, and the tests of nonlocality.  The experiments
can be analysed by making assumptions about detection efficiency,
and using the original inequalities, but these assumptions may
not be justified, leaving a loophole.  This is the detection loophole.

Since nonlocality is unique to quantum measurement, and appears
nowhere else in physics, we cannot afford to leave such a loophole.
We need to close it by improving the experiments and analysing them
unambiguously.  In some special cases, as suggested by Clauser, Horne,
Shimony and Holt, the real and ideal experiments are the same, and so
are the corresponding inequalities [9,10].  In other cases a real
experiment has critical inequalities that apply to all the outputs,
including those that are absent in the ideal experiment which it
simulates.  Weak nonlocality follows unambiguously from the violation
of any one of these critical inequalities.  They are almost always
different from the original inequalities for the ideal experiment, and
sometimes much more complicated.  These critical inequalities have
been obtained for particular experiments, for example [11,12], but in
general are relatively difficult to find.

This letter briefly describes the method used to obtain the critical
inequalities, and a computer program which provides them, for any Bell
experiment with two spatially separated subsystems.  A more general
and detailed description and program will appear in a later
paper. Zukowski et al.  [13,14] and more recently Pitowski and Svozil
[15] have proposed similar programs and applied them to special
cases.  Our program also provides an unambiguous test of weak
nonlocality for the results of all such experiments with adequate
statistics.  This part of the program does not use all the critical
inequalities, and so can be much faster than the one which does.

Some of the methods used here were developed for special cases by
Wigner [16], Garg and Mermin [11,12].  The general mathematical theory
was first developed by Froissart [17] and then Pitowsky [18].  A general
account of the relevant theory of input-output systems is given in
[19].

\vs
{\bf Nonlocality}

There is a profound distinction between experiments to test weak
nonlocality by the violation of Bell inequalities and most other
experiments on quantum systems.

Consider for example an experiment to determine the spectrum of an
atom, or a differential cross section for the scattering of an
electron by a molecule, or an experiment to determine the band gaps of
a solid, or to find a new particle.  The aim of all these experiments
is to determine the properties of quantum systems.  The classical
apparatus used to prepare the system and to make the necessary
measurements is essential, but secondary to obtaining these quantum
properties.

In Bell experiments the converse is true.  The aim is to test for
violation of the inequalities, which are derived from the
(statistical) properties of classical events, such as the setting of
the apparatus, which is a classical input, or the detection of a
particle by an electron avalanche, which is a classical output.  The
probabilities of the outputs, given the inputs, are what appear in the
Bell inequalities, and it is the location of these events in spacetime
that determine the locality or nonlocality.  The classical events are
connected by an ancillary quantum system, whose function is to produce
the unusual statistical properties of these classical events.  The
quantum properties, like the entanglement of particles, or the
polarization of photons, or the spins of atoms, are essential, but
secondary.  The primary result is the violation of an inequality,
which depends on the properties of the classical inputs and outputs.

We will only be concerned with those systems which have a finite
number of possible inputs and a finite number of possible outputs, so
we may use integers to label them, and there is also only a finite
number of the transfer functions $F$ described below.

Suppose a deterministic system consists of two subsystems $A,B$, with
corresponding inputs $\alpha,\beta$ and outputs $a,b$.  They are
related by a transfer function $F$ such that

\bl 
(a,b) = F(\alpha,\beta). 
\ee 
 
If the subsystems do not interact, then we expect $a$ to be a function
of $\alpha$ alone, independent of $\beta$, and $b$ to be independent
of $\alpha$, giving the separate relations

\bl 
a = F_A(\alpha),\h10 b = F_B(\beta), 
\ee 

which is a very special case.  If the output event $b$ is outside the
forward lightcone of the input event $\alpha$ and the output event $a$
is outside the forward lightcone of the input event $\beta$, this
independence is an inevitable consequence of special relativity, for
otherwise there would be signals faster than the velocity of
light. For this special case, $F$ is called a {\it local} transfer
function, and the remainder are {\it nonlocal}.  Deterministic
nonlocal transfer functions are never seen.

For stochastic systems, the relation between the inputs and outputs
can be expressed in terms of the measurable transition or conditional
probabilities 

\bl 
\Pr(\alpha,\beta\to a,b) = \Pr(a,b|\alpha,\beta).  
\ee 

The input-output properties of the system can also be expressed in
terms of the possible transfer functions $F$.  We can suppose that the
stochastic system sometimes behaves like a deterministic system with
one transfer function, and sometimes like another.  The probability
that the stochastic system behaves like a deterministic system with
transfer function $F$ is the {\it transfer probability} $\Pr(F)$.  The
transition probabilities are then

\bl 
\Pr(\alpha,\beta\to a,b) = \sum_F\Pr(F)
\delta\big((a,b),F(\alpha,\beta)\big),
\ee

where the delta function is equal to 1 if equations (1) are satisfied
for $(a,b)$ and $F(\alpha,\beta)$ and equal to zero otherwise.  The
stochasticity may be produced entirely by classical fluctuations, in
which case the sum on the right needs no nonlocal transfer functions.
Their probabilities can always be put to zero, as a consequence of
classical locality.

\vs
{\bf General Bell inequalities}

According to quantum dynamics and the quantum theory of measurement,
there could be systems for which it is impossible
to express the transition probabilities in terms of local transfer
functions alone.  At least one nonlocal transfer function may have to
have a nonzero probability $\Pr(F)$.  This is weak nonlocality.  This
definition of weak nonlocality in terms of transfer functions is
equivalent to Bell's original definition in terms of hidden variables
[19,1].

A general Bell experiment is an experiment to test this weak
nonlocality for experimental transition probabilities.
The general Bell inequalities are inequalities relating transition
probabilities, which follow from equation (4) when only local transfer
functions are included in the sum on the right hand side.  The
transfer probabilities themselves have to sum to unity and lie in the
range

\bl 
0 \le \Pr(F) \le 1.  
\ee 

When the sum over transfer probabilities is unconstrained, these
inequalities put no constraints on the transition probabilities other
than the obvious ones that they must also sum to unity for each input,
and lie in this range.  But if the sum is constrained to include only
local transfer functions, there are additional constraints on the
transition probabilities.  These are the Bell inequalities.  All Bell
inequalities are of this type.

For simple cases, like the original ideal Bell experiment, there is
little difficulty in deriving the inequalities, but as the numbers of
inputs and outputs increases, the number of inequalities, and the
difficulty in deriving them, increase very rapidly.  Since real
experiments have more outputs than the ideal ones, this is a practical
theoretical problem which has to be solved before the data can be
analysed for an unambiguous test of weak nonlocality.

Hence the need for a general computer program that derives the
critical inequalities effectively, and provides a test of weak
nonlocality for a given experimental set of transition probabilities.
\vs

{\bf Inputs and outputs}

To apply the theory to experiment, we have to know the inputs and
outputs.  For each run of an experiment, they are all classical
events.  The classical inputs are the settings of the devices that
determine what quantum variable is to be measured.  For example, when
the Paris group of Aspect and his collaborators [20] measured the
polarization of a photon, it was the setting of the effective angle of
orientation of the polarizers.  When the Geneva group of Tittel
measured the delay of a photon, it was the time difference in passing
through the long or short arm of an interferometer [21].  When Fry et al.
in Texas come to measure the spin of their atoms, it will be the
polarization angle of the laser that excites each atom [8].  In each
case, only a finite number of possible angles are used, and these can
be labelled by an integer.

In all experiments to date, the classical outputs occur in particle
detectors.  Suppose there are $\N(A,d)$ detectors in subsystem $A$ and
$\N(B,d)$ in $B$.  For each run of the experiment, a given detector
may or may not fire, so altogether the number of possible outputs for
the whole system is

\bl
\N(a)\N(b) = 2^{\N(A,d)+\N(B,d)}.
\ee

In practice some of the outputs for the whole system may have negligible
probability, in which case they could be ignored, and the effective
number of outputs reduced.  But this simplification complicates the
use of the computer program and is not recommended.

To use the program, first establish the notation.  For each subsystem,
label the detectors by an integer $d$, where

\bl
d=0,\dots \N(A,d)-1\hx    {for subsystem $A$, and}
\ee\bl 
d=0,\dots \N(B,d)-1\hx    {for subsystem $B$.} 
\ee

For one detector the output is a binary digit $f$, which is equal to
one if the detector fires in a given run, and zero otherwise.  For a
single
rune we denote this output for the detector labelled $d$ of subsystem $A$
by
$f(A,d)$, and similarly for $B$.  The output for the subsystem
$A$ or $B$ is represented by the integer

\bl
a = \sum_{d=0}^{\N(A,d)-1} f(A,d)2^d\hx{or}
\ee\bl
b = \sum_{d=0}^{\N(B,d)-1} f(B,d)2^d
\ee

and the total number of possible outputs for each subsystem is

\bl
\N(a) = 2^{\N(A,d)} \hx{or} \h10\N(b) = 2^{\N(B,d)}.
\ee

If required, the output $(a,b)$ for the whole system can be represented
by the single integer $\N(b)a+b$.

\vs

{\bf The experiment and the computer program}

Bell experiments test Bell inequalities or one of its generalizations.
Every Bell experiment has classical inputs and outputs.  The inputs are
the
settings of the apparatus that determine the measurements that are to
be made.  Typically these settings are orientations of polarizers.
The outputs are the detections of particles.  Typically these are
electron avalanches produced by incident photons, or possibly detection
of ions or electrons.  

The settings must be sufficiently late in time, and the detections
sufficiently early, that there are limits on the possible propagation
of signals between them.  The inputs and outputs belong to a finite
number $\N(s)$ of subsystems, such that for any one of these
subsystems, no signal can go from {\it any} input or output of one
subsystem to {\it any} input or output of another.  For example in the
original Bell experiment [1], or the CHSH experiment [9], $\N(s)=2$, for
entanglement-swapping $\N(s)=3$, and for
Greenberger-Horne-Shimony-Zeilinger (GHZ) experiments either $\N(s)=3$
or $\N(s)=4$ [22].  For the ideal GHZ experiments, weak nonlocality
follows from equalities, but for the real experiments, these become
inequalities.

In general, experiments that test weak nonlocality by means of Bell 
inequalities can be classified by a string of positive integers:
$$
(N(s),N(0,d),N(1,d),.....,N(N(s)-1,d),N(i))
$$
where $\N(s)$ is the total number of subsystems, $\N(k,d)$ is the
maximum number of detectors in subsystem $k$, and $\N(i)$ is
the total number of settings, which we assume to be the same for all
subsystems. In the first edition the program can be used to analyse
only those experiments for which $\N(s)=2$ and therefore the integer
triplet \bl (N(A,d),N(B,d),N(i)) \ee characterizes any such setup. The
generalization to an arbitrary number of subsystems, which will be
required for later editions, is straightforward.

BellTest is a computer program. We will describe the first edition which
performs two independent functions: (i) 
it provides a test for the violation of locality using raw experimental
data, 
and (ii) it produces all the Bell inequalities for a given experimental 
setup.

The program operates in the following manner: For a given integer 
triplet of the form (12), the set of all possible inputs and outputs for
this 
class of experiments is generated. Although the program is restricted to
cases 
involving equal numbers of inputs, in general $\N(A,d)\not=\N(B,d)$ and
can
therefore handle unequal numbers of outputs. A particular experiment is 
defined by specifying the set of settings of the classical measurement 
apparatus. The program then proceeds to carry out functions (i) and (ii).

For function (i) one must specify a set of conditional probabilities,
obtained from the raw 
experimental data, that may or may not satisfy the corresponding Bell 
inequalities. Nothing in the raw data must be neglected,
failing to do so results in biased statistics. This test amounts to the
linear programming problem of checking for the possibility to 
satisfy a given set of constraints. In our case the quantities to be
varied 
are the transfer probabilities (5), subject to the equalities that the
linear 
combination must be the same as the given conditional probability, their
sum 
must be unity, and that every transfer probability must be greater than
zero. 
The linear programming analysis is done using a freely available software 
called LP{\_}Solve [23].

Mathematically, such a space is constrained to exist inside a 
unit-hypercube; that being the generalization of the ordinary cube to 
dimensions greater than 3, and with vertices whose coordinates are
composed of
0/1 entries only. Geometrically, the conditional probability space is a
bounded
convex polyhedron, otherwise known as a polytope, which is  given as the
convex
hull of its vertices. Each vertex has coordinates that are strings of 0's
and 
1's and represent transfer functions of deterministic systems. For a
general 
account of polytope theory the reader is advised to read Ziegler's
excellent 
text on the topic [24].

Polytopes can be described in terms of their vertices
($V$-representation),
or in terms of equations and facet-defining inequalities
($H$-representation),
where a facet is a surface of dimension one-less than that of the polytope 
itself. The generalized Bell inequalities define such surfaces. The 
problem of transforming from the $V$-representation to the
$H$-representation 
is called {\it facet enumeration} and the reverse of this is 
know as {\it vertex enumeration}. We are interested in the former. 
Computational geometers have developed many tools for the analysis of 
polytopes, and we make use of a software called Polyhedron Representation 
Transformation Algorithm (PORTA) [25] to generate all the Bell 
inequalities for a given experimental setup. However, polytopes of this
type 
are infested with large numbers of facets, and as a result it may take a
long 
time to enumerate them all [26].

\vs
{\bf An Example}

Let us consider an example so as to put BellTest into perspective. Suppose
that our experiment is of the class (2,2,2). The integer triplet has
$N(i)=2$, which defines two possible apparatus settings for each
subsystem. These are labeled 0 and 1. In general, for $N(i)$ settings  we
define the following integer set of possible apparatus settings
${0,1,....,N(i)-1}$. Each subsystem is assigned the same set of integers.
However the set of integers for subsystem $A$ do not necessarily define
the same set of apparatus settings as defined by the integer set assigned
to subsystem $B$. For example, Clauser, Horne, Shimony and Holt [9]
proposed an experiment where each adjustable apparatus has two possible
settings, that is subsystem $A$ has settings $a$ and $a'$, and subsystem
$B$ has settings $b$ and $b'$. With BellTest, subsystem $A$ has settings
0, representing $a$, and 1, representing $a'$. Subsystem $B$ also has
settings 0 and 1 but here they represent $b$ and $b'$ respectively. For
the system as a whole, that is $A+B$, there are four inputs
$(\alpha,\beta)$, which define the set of possible experiments that we may
carry out.
$$
(0,0) (0,1)
$$
$$
(1,0) (1,1)
$$

Suppose that in our experiment the adjustable apparatus of each subsystem
is prepared such that we only investigate setups (0,0), (0,1), and (1,1).
For completion, let us suppose that $\alpha$ and $\beta$ are the
orientations of a polarizer, perhaps a calcite crystal. Each subsystem has
two detectors, each of which has two states, triggered (labelled 1) and
not triggered (labelled 0), that record the emergence or non-emergence of
a photon in the ordinary or extraordinary ray. All possible combinations
of detector events for each subsystem define an output. For the case at
hand each subsystem has four possible outputs, which we label 0,1,2, and
3, and therefore, the whole system has 16 possible outputs $(a,b)$.

For this example, the space of all conditional probabilities is given by
the convex hull of 256 transfer functions. Using PORTA, BellTest lists 48
inequalities, some of which are of the Bell type.

Given raw data in mode (i), BellTest checks if a given set of conditional
probabilities satisfy the corresponding Bell inequalities.  

\vs
{\bf Closing Remarks}

In this letter we have described a tool to test for violations of the
critical Bell inequalities and 
therefore the closure of the detection loophole for any Bell
experiment with two subsystems. The analysis involves making use of the
$entire$ raw data set. In 
addition, the user can use BellTest to list all the Bell inequalities if 
desired, but it must be stressed that the number of such inequalities
grows 
very rapidly with respect to the number of transfer functions representing 
deterministic systems for a particular class of experiments (12). We
invite 
anyone who is interested in using BellTest to download the first edition
from 

\cl{http://www.strings.ph.qmw.ac.uk/QI/main.htm}.

\vs
{\bf Acknowledgements}

We thank Ed Fry, Nicolas Gisin, Serge Haroche, Monique Laurent, Boris
Tsirelson
, Thomas Walther and Ting Yu for helpful and stimulating communications
and the
Leverhulme Trust, the European Science Foundation and PPARC for
support.

\vs{\bf References}

[1] J.S. Bell, Physics 1 (1964) 195.

[2] I.C. Percival, Phys. Lett. A 244 (1998) 495.

[3] P. Pearle, Phys. Rev. D 2,(1970) 1418.

[4] E. Santos, Phys. Rev. A 46 (1992) 3646.

[5] P.G. Kwiat, P.H. Eberhard, A.M. Steinberg, and R.Y. Chiao, Phys. Rev.
A 49 (1994) 3209.

[6] E.S. Fry, T. Walther and S. Li, Phys. Rev. A 52 (1995) 4381.

[7] N. Gisin, and B. Gisin, Phys. Lett. A (1999) 323.

[8] E.S. Fry and T. Walther, in Adv. Atom. Molec. and Opt. Phys. 42 
(Academic Press, New York, 2000) 1.

[9] J.F. Clauser, M.A. Horne, A. Shimony, R.A. Holt, Phys. Rev. Lett. 23
(1969) 880.

[10] J.F. Clause, M.A. Horne, Phys. Rev. D 10 (1974) 526.

[11] A. Garg and N.D. Mermin, Found. Phys. 14 (1984) 1.

[12] A. Garg and N.D. Mermin, Phys. Rev. D 35 (1987) 3831.

[13] M. Zukowski, D. Kaszlikowski, A. Baturo, J. A. Larsson,
quant-ph/9910058.

[14] D. Kaszlikowski, P. Gnacinski, M. Zukowski, W. Miklaszewski and A.
Zeilinger, quant-ph/0005028.

[15] I. Pitowski and K. Svozil, quant-ph/0011060.

[16] E. Wigner, Am. J. Phys. 38 (1970) 1005.

[17] M. Froissart, Nuovo Cimento B 64, 241 (1981).

[18] I. Pitowsky, Quantum Probability - Quantum Logic. Lecture Notes in
Physics 321 (Springer-Verlag, Berlin, 1989).

[19] I.C. Percival, quant-ph/9906005.

[20] A. Aspect, J. Dalibard, and G. Roger, Phys. Rev. Lett. 49 (1982)
1804.

[21] W. Tittel, J. Brendel, H. Gisin, and H. Zbinden, Phys. Rev. A 59
(1999)
4150.

[22] D.M. Greenberger, M.A. Horne and A. Zeilinger, in Bell's Theorem,
Quantum Theory and Conceptions of the Universe, ed. M. Kafatos (Kluwer
Academic Publishers, 1989) 69.

[23] M. Berkelaar, LP{\_}Solve Version 3,
ftp://ftp.es.ele.tue.nl/pub/lp{\_}solve/.

[24] G.M. Ziegler, Lectures on Polytopes, Graduate Texts in Mathematics
152 (Springer-Verlag, New York, Berlin Heidelberg, 1995).

{\obeylines
[25] T. Christof and A. Lobel, PORTA,{http://elib.zib.de/pub/Packages/
mathprog/polyth/porta/}}.

{\obeylines
[26] G.M. Ziegler, Lectures on 0/1-Polytopes,
http://www.math.tu-berlin.de/\~{}ziegler/}.

\end{document}